\documentclass{sig-alternate}
\usepackage{amsfonts}
\usepackage{amscd}
\usepackage{amsmath}
\usepackage{subfigure}
\usepackage{graphicx}
\usepackage{longtable}

\usepackage{times}

\usepackage{url}
\usepackage{boxedminipage}
\usepackage{ifthen}
\usepackage{floatflt}

\newcommand{\figref}[1]{Figure~\protect\ref{#1.fig}}

\newboolean{eps}
\setboolean{eps}{false}

\begin{document}

%\conferenceinfo{CCS'10,} {October 4--8, 2010, Chicago, Illinois, USA.}
%
%\CopyrightYear{2010}
%
%\crdata{978-1-4503-0244-9/10/10}
%
%\clubpenalty=10000
%
%\widowpenalty = 10000

\title{{Phishing, Personality Traits and Facebook}}

%\author{
%Tzipora Halevi\inst{1} \and James Lewis\inst{2} \and Nasir Memon\inst{3} \and Tuo Xiang\inst{2}
%}
%
%\institute{Polytechnic Institute of New York University\and University of Michigan-Dearborn
%\and University of Alabama, Birmingham 
%}

\numberofauthors{3}
\author{
\alignauthor
Tzipora Halevi\\
       \affaddr{Electrical and Computer Engineering}\\
	\affaddr{Polytechnic Institute of New York University}\\
       \affaddr{Six MetroTech Center}\\
       \affaddr{Brooklyn, NY 11201}\\
       \email{thalev01@students.poly.edu}
% 2nd. author
\alignauthor
James Lewis\\
       \affaddr{Technology Culture and Society}\\
	\affaddr{Polytechnic Institute of New York University}\\
       \affaddr{Six MetroTech Center}\\
       \affaddr{Brooklyn, NY 11201}\\
       \email{JLewis@Poly.edu}
       \alignauthor
Nasir Memon\\
       \affaddr{Computer Science}\\
	\affaddr{Polytechnic Institute of New York University}\\
       \affaddr{Six MetroTech Center}\\
       \affaddr{Brooklyn, NY 11201}\\
       \email{memon@nyu.edu}
}
%%Tzipora Halevi\\
%%{Electrical and Computer Engineering}\\
%%{Polytechnic Institute of New York University}\\
%%{\texttt{thalev01@students.poly.edu}} \and
%%Nitesh Saxena\\
%%{Computer Science and Engineering}\\
%%{Polytechnic Institute of New York University}\\
%%{\texttt{nsaxena@poly.edu}}
%%}

\date{}

\maketitle

\begin{abstract}

%TODO: add abstract
Phishing attacks have become an increasing threat to online users. 
%TH 9/18/1
%Little research has been conducted into the factors that cause people to respond to them. 
Recent research has begun to focus on the factors that cause people to respond to them.
Our study examines the correlation between the Big Five personality traits and email phishing response. We also examine how these factors affect users behavior on Facebook, including posting personal information and choosing Facebook privacy settings.

Our research shows that when using a prize phishing email, we find a strong correlation between gender and the response to the phishing email. In addition, we find that the neuroticism is the factor most correlated to responding to this email. 
Our study also found that people who score high on the openness factor tend to both post more information on Facebook as well as have less strict privacy settings, which may cause them to be susceptible to privacy attacks. 
%TH 9/19/12
In addition, our work
%Our work also 
detected no correlation between the participants estimate of being vulnerable to phishing attacks and actually being phished, which suggests susceptibility to phishing is not due to lack of awareness of the phishing risks and that real-time response to phishing is hard to predict in advance by online users.

%is not due to the fact that users not being aware of phishing risks 
%and is hard to predict by users.
%to phishing is hard to predict, and 
% and the likelihood of falling for an email phishing attack.

%The increasing popularity of Online Social Network sites, such as Facebook, and the information sharing by its users have privacy implications which have yet to be fully understood. We examine how the behavior of users on Facebook is related to their Big-Five personality traits and the way they affect the likelihood of falling for an email phishing attack. 

We believe that better understanding of the traits which contribute to online vulnerability can help develop methods for increasing users' privacy and security in the future. 
%Specifically, defenses should be developed that concentrate on people that may test high on neuroticism 

\end{abstract}

% A category with the (minimum) three required fields
%\category{D.4.6}{Security and Protection}{Authentication}
%\category{H.4}{Information Systems Applications}{Miscellaneous}
%\category{C.2.0}{Computer-Communication Networks}{General}
%\category{D.2.8}{Software Engineering}{Metrics}[complexity measures, performance measures]
\category{H.5.m.}{Information Interfaces and Presentation (e.g. HCI)}{Miscellaneous}

\terms{Security, Human Factors}

% NOT required for Proceedings

%\newpage

%\vspace{-3mm}
\section{Introduction}
\label{sec:intro}
%\vspace{-3mm}

With the increased popularity of the internet, people spend more time online. Among the more popular online activities are email communication as well as participating in social networks, such as Facebook. As a result, email attacks and privacy threats pose increasing security concerns for online users.

One such threat is phishing email attacks. These attacks
%Phishing attacks 
attempt to acquire personal information, such as user-name and passwords, through fraudulent emails and represent a form of social engineering techniques used to deceive users. Phishing attacks have been on the rise in the last few years, with phishing emails becoming more targeted, using personal information about their intended victims, in an attempt to seem like authentic emails and improve the response rate to the attacks.
%towards their intended victims. 

In this work, we set out to investigate the connection between Phishing vulnerability, personality traits and Facebook activity. For this purpose, we present a study that examines how psychological traits correlate to deception detection and phishing response. 
%We explore the vulnerabilities of people to online threats and the correlation to different personality aspects. 
%Specifically, we examine the response to phishing attacks and the tendency to post personal information on Facebook and how it relates to certain psychological traits. Identifying the personality types that may make some people more vulnerable to online threats is an important step in creating defenses and protecting users from online privacy vulnerability and email attacks.
%Another aspect we plan to examine relates to phishing attacks. 
%Phishing attacks are a way of attempting to get personal information from users and represent a form of social engineering techniques used to deceive users. Previous research showed a responserate of up to 70\% for phishing attacks on social networks.  
%We look at the connection between deception detection and personality traits, and how it affects the likelihood of falling for an email phishing attack. 
We also examine the tendency to post personal information on Facebook and how it relates to certain psychological traits as well as responding to phishing emails.

Our work follows the hypothesis that responding to phishing emails represents an error in judgment, which is due to certain emotional biases. The ability to provoke such emotional triggers may be connected to the specific personality traits, where people who score high on certain traits may be more likely to fall victims to such attacks. 

Further, the ways in which personality traits manifest themselves in off-line behavior could have a similar affect on online behavior as well. Previous studies linked neuroticism to the tendency to believe people (and failure to detect lies).
%TH 9/19/12
Premeditation was linked to the
%and reduced
ability to point to suspicious scam messages (when examined off-line), which may affect vulnerability to online phishing scams as well.

Despite the rise in phishing attacks, their connection to psychological factors and to social networks behavior has not been thoroughly explored. Identifying the personality 
%traits)
characteristics that may cause higher vulnerability to online threats is an important step in creating defenses and protecting users from email attacks and online privacy threats. 
%We conduct a study with over 100 users to determine these correlations and how they can be used to help predict privacy-related online behavior. We believe that better understanding of the reasons for online vulnerability can help prevent such behavior and increase users privacy and security in the future.

%\newpage
\subsection{Scams and Personality}
In classical decision theory, decision making under risk is assumed to be based on pure logic. Under these assumptions, reasonable people make rational choices based on objective factors. However, Kahneman et al. \cite{KT79} have shown that people's decisions tend to be biased and are not purely logical.

A scam is a pretense in which a fraudulent attacker attempts to extract valuable information or monetary gain from the victim. A response to scam can be viewed as a decision error, where the user does not estimate correctly the risk, due to certain biases.

The popularity of different scams is due to the fact that a certain percentage of people tend to fall for them. They provide the malicious attacker with an opportunity to steal 
%TH 9/17/12
the victim's 
personal information.
%for personal gain. 
In addition, many scams attempt to get money directly from the scam victims.  

Scams appeal to different human vulnerabilities, such as the desire for immediate gain, the desire to help people (which causes African scams to be successful) and the desire to be liked by the scam initiators. It has been suggested that certain people have ``victim personalities'' that make them more vulnerable to scams. These victims may fall for scams repeatedly.

Studies of the psychology of scams show that victims often respond to emotional triggers. These triggers include greed, fear, heroism and desire to be liked.  People also tend to obey authority, and scams which use authoritative words (such ``official'') are more likely to get response. Another factor (which is also used in traditional marketing) is making an opportunity seem scarce, or getting the scam victim to feel he made a commitment, by responding to the scam offer.

%Certain types of scams appeal to different human emotions. For example, African scams tend to appeal to greed, lust or heroism. 

One of the factors that may 
%TH 9/17/12
make it more likely for
%cause 
certain people 
%TH 9/7/12
%to be more likely 
to become victims is the lack of emotional control. A research by the University of Exeter \cite{ScamPsychology}, examined the reasons for the lapses of judgment by scam victims and was based on interviews and questionnaires filled by scam victims and examination of current scams.
It found that scam victims reported being unable to resist responding to persuasion and being undiscriminating about the offers they respond to. The research suggests people who are socially isolated may be more vulnerable to responding to scams.
Victims response also indicated that some of the people viewed responding to the scam as taking a gamble, 
%TH 9/17/12
where
%viewing 
their initial investment in the scam 
is
%as 
small in comparison to the larger prize. 
%The research suggested 
One of the study conclusions was
that there is a particular segment of people (about 10-20\% percent of the population) who are particularly vulnerable to scams. Some people become serial scam victims, who fall repeatedly for scams. 

In \cite{LS01}, Langenderfer et al. identified the fact that scam messages often attempt to present a unique opportunity and require urgent response. These techniques are used in legitimate sales and marketing as well and are believed to be effective.  

One of the defenses against scams is consumer education. However, since scams continue to change, educating the population about existing scams will have limited effect. This can be evidenced by spear phishing attacks, which are a new generation of email targeted attack. These attacks are more sophisticated than traditional phishing attacks and are harder to detect by the users. 
%many scams(which African scams belong to) attempt at invoking strong emotions such as greed, guilt, lust or heroism \cite{CNC07}.

Research into lie detection 
%TH 9/19/12
by Enos et al. 
\cite{EBCG06} also found that people who scored high on neuroticism had a significantly worse probability of detecting lies and had a tendency to overestimate the level of truth in other people's responses. neuroticism may cause people to be more upset when being lied to and therefore cause people to prefer believing that other people are generally truthful. This may indicate that people with a high level of neuroticism may be more vulnerable to scams in general. On the other hand, agreeableness was positively correlated to successful lie detection. This may indicate that people with this trait are more compassionate and sensitive to other people's responses.

%TH 9/19/12
%In \cite{EBCG06}, Enos et al. compared between machines and human ability to detect human lies. They examined the correlation between different personality traits and detection probability. They found that neuroticism was inversely proportional to the ability to detect lies. The authors speculated that people who scored high on neuroticism find lies more upsetting and therefore tend to believe that people are generally truthful. 

%This may indicate that people with a high level of neuroticism may be more vulnerable to scams in general.
%is a main factor in human speech and scamming ability. In this case, a scamming person which has high neuroticism value may be able to better deceive his victims.

The relationship between personality and scam victims has been further explored in \cite{ML12}. In this paper, people were shown different offers and were asked to identify which ones were scams. Scam victims were identified as having certain personality traits. Specifically, premeditation (which is part of the impulsivity test) was highly correlated to avoiding scams. Extroversion also was detected as a predictor to avoiding scams. Introvert people may be more likely to fall to scams due to their preference for internet communication (over face-to-face communication), which is a medium highly exploited for scams. Also, the paper speculates that less human contact may make Introvert people less familiar with negative experiences of other scam victims. 

However, research is divided on the contribution of some personality traits to scams. For example, while some work showed that people who are agreeable are better equipped to detect lies \cite{EBCG06}, in other scenario agreeable people were found to be more likely to fall for scams \cite{ML12}. 

\subsection{Personality Types and Internet Behavior}
%TH 9/14/12
Research into cyber-security has begun to look at how different aspects of psychology can affect the end user and therefore compromise Internet security.
One existing concern is that the internet may replace normal social activities and that people who are preoccupied with the internet may be compensating for loneliness and social seclusion.

%In \cite{ZL11}, 
A study by Zhou et al \cite{ZL11} examined the user acceptance of mobile commerce and found that neuroticism had a negative effect on its perceived usefulness. In contrast,
research by Wolfradt et al. \cite{WD01} found a high interest in using the internet for communication purposes in people who scored high on the neuroticism scale.

%TH 9/5/12
A few studies found gender-based differences in online activity. Milne et al \cite{MLC09} of online shopping services found that male were more likely to engage in risky online activity. On the other hand, Byrne et al.  
found that women were more likely than men to click on a link with a coupon even when being warned of a potential threat.

%In two 
Two 
studies by Hamburger et al. \cite{AB03, HB00}, 
which explored
the personality of heavy internet users,
% was explored.
also detected differences between the genders.  
%These studies found differences between the genders. 
In particular, 
their research showed that 
for women, neuroticism was positively related to loneliness, while for men, the correlation was significantly lower. Also, for women, both neuroticism and the feeling of loneliness where positively related to the use of social services, while extraversion was negatively related to both. For men, the correlation was significantly lower to neuroticism (and was uncorrelated to loneliness).
One explanation for these results may be that women are more sensitive to their emotional and social needs and 
%open to accept
realize
the ability of the internet to help fill those needs.  

In another research, Schrammel et al. \cite{SKT09} examined if there is a relationship between personality traits and disclosure of information online but did not find any correlation between them. However, the study found that people who spend more time online provide more information on their profile.

% or use it heavily do it to replace missing social contacts The personality of internet users and its effect on their online behavior

%such as getting the victim to look forward to a future positive experience when the receiver gets his prize or product. People also tend to respond to authority, 
%Certain personality traits may make some people more vulnerable to respond to such emotional triggers.

%\vspace{-3mm}

%methods to raise awareness and detection ability of such emails.
% is imperative to combating these attacks 
%and developing defenses which respond to those vulnerabilities.

%
\subsection{Phishing Vulnerability} 
%\subsection{Phishing Vulnerability and Personality Traits}
%\vspace{-2.5mm}

%In this work, we set out to investigate the correlation between Phishing vulnerability, personality traits and Facebook activity.

Phishing is an attack that uses fraudulent electronic mail (email) that claims to be from a trustworthy source.
The goal of phishing emails is to get personal information from the users, such as user ID and passwords. The attacker can than use this information to impersonate a user and access the user account for financial gain. 
In the last few years there have been a significant increase in Phishing and Spear phishing activity, with many of the emails designed to target directly their victims in an effort to raise the likelihood that the user will respond to the emails.

The direct damage of phishing 
is due to
%includes 
the costs of goods or money stolen,
%money or goods stolen, 
which has been estimated to be over 1 billion dollars \cite{JM06}. However, the damage 
%TH 9/19/12
also
%of phishing 
includes overhead to companies that get attacked, such as customer service support needed to respond to user calls. In addition, as people become aware of the dangers of phishing, they may avoid performing online purchases and online banking, which results in reduced business to companies who offer their services online.

Previous studies of phishing looked into the technical understanding (or the lack of it) that makes people fall for phishing and for methods to improve the user ability to detect such attacks.

%In a study by 
Dhamija et al. \cite{DTH06}
explored
the reasons that people respond to phishing attacks.
% were explored. 
Test participants were shown 20 websites and were asked to determine which ones were likely to be authentic and which were not. The study found that many of the user were not familiar with the technical cues of secure websites. 
%TH 9/19/12
Those
%Many of the 
users either did not examine the address bar or the status bar, did not look for "https" at the beginning of the website address nor looked for the padlock sign. This implies that standard security indicators may not be useful in many cases as users do not understand them or neglect to search for them, even when actively trying to determine if a site is authentic. 

% and looked for ways to improve the user 
One of the suggested defenses for phishing is increased education for internet users. However,
research into phishing vulnerability \cite{CAP11} shows that education has limited effect on phishing response. In this study, Caputo et al. sent three separate phishing emails to all workers in a medium-sized company. Each email was sent three months after the previous one. Training sessions were conducted in-between to raise the people awareness about the dangers and signs of phishing emails. The study found the training had a limited effect, where over 30\% of the participants clicked on each of the emails and 10\% of the respondents clicked on all three phishing emails. The study also found that 7\% of the participants did not click on any of the emails (demonstrating that some people are less susceptible than others to fall for phishing attacks).

Sheng et al. \cite {SHK10} performed a demographic study of phishing susceptibility. Their study found that women were more likely to fall for phishing ($53\%$ of women and $41\%$ of the men fell for the phishing experiment). The women in the study had less technical expertise, which may account for some of the difference in phishing response. However, the women did have a higher level of familiarity with anti-phishing education, which further supports the hypothesis that anti-phishing education may not be a significant factor in phishing prevention.

%TH 9/14/12
%Research on cyber-security has begun to look at how different aspects of psychology can affect the end user and therefore compromise Internet security. It has been theorized that the five factor model would be most likely useful in detecting people that are more likely to be susceptible to phishing attacks.

%TH 9/14/12 Chuchuen study methodology and results are not clear
%Chuchuen et al. explored the relationship between the DISK personality model and the ability to potentially recognize suspicious phishing links by asking test participants questions about different phishing techniques as well as about their personality.

Our research assumes that responding to phishing, just like responding to scams, results from an error of judgment. Our goal is to understanding the psychological traits that cause certain people to be make such errors. In addition, we seek to see if these correlate to other lapses of judgment in online behavior (such as posting data on social networks sites).

The success of a phishing attack depends on users responding to it and providing their information. Therefore, understanding the psychological 
reasons for responding to such emails is imperative to 
developing
%and finding 
effective defenses against such phishing attacks.

\subsection{Facebook Privacy} 

%TH 9/5/12
%The increasing popularity of Online Social Network sites, such as Facebook, and the information sharing by its users have privacy implications that are yet to be understood. Research studies and known examples demonstrate the fact that people under-estimate the risks in sharing information online. Specifically, users share too much information, Facebook does not adequately protect user privacy and third-parties actively seek information about Facebook users. Personality traits are  believed to influence the use of social media and also have an effect on Internet security awareness. 

Facebook has become the most popular social networking site, with over 900 million users to date. The application allows people to post text messages, share photos and put other personal information online, such as birth 
%TH 9/19/12
date, address, work place and other data. Users have lists of friends who can also post messages on their site. This results in a large amount of personal information shared between many users. While privacy settings can be changed on Facebook, many users leave their information public to all Facebook users or may set them open to viewing by friends and their friends. Since the average friends list has 190 people, this results in sharing the information with a large number of people. Since people may also tag other people (who appear in their pictures), a person's private information may also be leaked by other Facebook users. Overall, Facebook sharing may result in privacy threats to Facebook users, who may not be fully aware of the implications of sharing personal and sensitive data. 

While studies into users' online privacy attitude have shown that most users are concerned with the way their data will be used \cite{ACR99, CRA99}, research and 
%9/5/12
%Research studies 
%and 
known examples demonstrate the fact that people under-estimate the risks in sharing information online. 
%Specifically, users tend to share too much information, 
Facebook does not adequately protect user privacy and third-parties actively seek information about Facebook users. 
%In a study by 
Privacy International \cite{PI07} 
%that 
examined 21 online service companies and assessed Facebook as 
one of seven sites that
%a site that may 
pose substantial privacy threats. 
%TH 9/14/12
Egelman et al. \cite{EOK11} showed that Facebook users tend to make mistakes when choosing their privacy settings, which were likely to result in sharing information with unintended parties. 
However, users tend to ignore these risks.
% and share too much information online.
In a study by Debatin et al., which included 119 students, Facebook users were found to perceive the benefits of sharing information on Facebook as significantly higher than the risks associated with sharing this information. This was further supported by 
%a study by 
Govani et al. \cite{GP05}, 
%which 
who
found that users may be willing to take higher security risks to enjoy the benefit of certain online services. This indicates that privacy threats may increase due to the fact that many users underestimate or ignore the privacy risks in sharing personal information while focusing on the advantages of the social network. 

Personality traits are believed to influence the use of social media and also have an effect on Internet security awareness. Our research examines how the traits affect Facebook-related decision making and behavior. Our goal is to detect the characteristics of users who may be more susceptible to privacy threats.

%cd cdcd and gain further understanding of the variables which need to be addressed to develop efficient  

\subsubsection{Personality Types and Facebook Use}

%There is a great deal of discussion as to how personality affects Facebook use

A few studies examined the relationship between personality traits and Facebook related behavior. 
%TH 9/15/12
Most research has focused on the hypothesis that real-world personality is most likely expressed in the cyber-world in a similar way.
In \cite{GAV11}, Gosling et al.
%the relationship between personality traits and Facebook related behavior was explored. The 
found that extraversion is related to the frequency of Facebook use and engagement with the site. This suggests that the users on-line personality is directly related to their off-line personality.
In another research, Qaurcia et al. \cite{QLS12} also found that the number of friends users have on Facebook was directly related to extraversion, while no significant relationship was found to other personality traits. 

\subsection{Big Five Framework}
%TH 9/14/12
Personality is a consistent pattern of how people respond to stimuli in 
%your 
their
environment and their attitude towards different events.
%Personality traits affect people's attitude towards different events. 
The five factor model of personality assessment is currently one of the most widely used multidimensional measures of personality \cite{MJ92}.
Its goal is to encapsulate personality into five distinct factors which allow a theoretical conceptualization of people's personality. These dimensions are Neuroticism, Extroversion, Openness, Agreeableness, and Conscientiousness. One of the most widely used measures of this five factor model was developed by Costa and McCrae and is called the NEO-PI FFM test \cite{CM92b}. This is a short 60 question test that allows for relatively quick, reliable, and accurate measurement of participants personality across these five major dimensions of personality.
%TH 9/12/12
%The Big Five framework from Costa \& McCrae provides a method of accurately evaluating the following main personality traits - Openness, Conscientiousness, Extraversion, Agreeableness and Neuroticism (also referred to by the ``OCEAN'' acronym) . Specifically, we are using the NEO-PI test \cite{CM92b} of that framework to evaluate these five traits based on user response to a questionnaire with 60 items.
%TH 9/12/12
%The Big Five framework of personality traits (known as OCEAN - Openness, Conscientiousness, Extraversion, Agreeableness and Neuroticism) from Costa \& McCrae provides a method of accurately evaluating these main personality traits based on user response to a questionnaire with 60 items.
%provides an instrument for accurately evaluating 
%has been identified as a robust model for understanding the relationship between personality and various academic behaviors. 
%TH 8/27/12
This model is considered superior to other models in capturing the common elements of personality traits and providing a precise personality structure description \cite{W04}. 

Studies demonstrate that the five factors manifest themselves in certain patterns of behavior, and are found in different age, gender and race groups. In addition, there is evidence that the traits are hereditary, which suggests an underlying biological basis \cite{CM92}.
% of personality functioning. 
%TH 9/14/12
%One of the benefits of the five factor model is that it has been useful in testing several different constructs, including employment \cite{RC03} and education.
The advantages of the model led to its integration in a wide array of previous personality traits-based studies 
in different fields, including employment \cite{RC03} and education \cite{BPE99}.
The 
%big five 
framework has 
%also 
been identified as a robust model for understanding the relationship between personality and various academic behaviors. 
Our research sets to examine if this relationship extends to online security and privacy-related behavior. 

%TH 8/27/12
Determining the personality factors that contribute to vulnerability to phishing attacks as well as privacy threats is an important step towards improving online security. This can help in creating customized defenses to improve user awareness and protect people who may be more vulnerable to such privacy and security attacks. 
%TH 8/27/12
%Our hypothesis is that certain personality traits are related to online behavior.
%The relationship between those traits and online behavior is important to understanding the potential sources of online vulnerability and what makes some people vulnerable to privacy threats as well as phishing attacks. 
%This can be further used to improve user awareness as well as explore those traits to create better defenses geared to protect people who may be more vulnerable to such privacy and security attacks.

%For example, we examine how neuroticism affects the way users set their privacy settings on Facebook and whether it is related to higher security awareness. On the other hand, we check if Extraversion and Openness predict how much information users will share online.

%\newpage

\subsection{Overview of Contributions}

In this work, we try to identify personality traits that cause higher vulnerability to phishing attacks. We examine the correlation to social networks activity and try to see if we can identify personality traits that may cause privacy threats.

This research is the first one we know of that correlates between phishing, personality traits and Facebook activity. We also examine the correlation to other factors, such as gender, general online usage characteristics and online pessimism.

Our research shows that certain personality traits are more likely to be associated with vulnerability to phishing attacks as well as with online information sharing on social networks site.

%\vspace{-3mm}
\smallskip

\section{Overview of Our Experiments}
\label{sec:setup}
%\vspace{-3mm}

\subsection{Methodology}

Participants were 100 students drawn from a psychology class at a small Northeastern engineering college. Students participated for extra credit and were told that this was primarily a study on Internet usage and beliefs. 
There were 83 males and 17 female.  
Students ranged from  18 to 31 with an average age of 21.17 years with two student choosing to to disclose their age. Students ranged in a variety of different majors but were primarily in the science and engineering disciplines

%We had 107 test participants. All test participants were Polytechnic Institute of NYU students who took a psychology-related class. The participants were both undergrad and grad students. The students were told this was a general internet usage study.

\subsection{Personal questionnaire, personality traits and Facebook activity}

In the first part of the experiment, the students were given a link to an online questionnaire and were asked to fill it within a week. The reason the questionnaire was put online was to prevent in-class interaction that may affect the results.

The questionnaire included three parts: A personal questions part, which included the users email, age, academic and work background information. It also included an online activity section. In this section, the users were first asked 
%TH 9/7/12
to assess as a 7-point scale (from 1 = not very likely to 7 = definitely) 
their online activity
%about their usage activity 
and their estimate of the probability of bad consequences happening 
to them
online. They were also asked about the types of data they put on their Facebook account, the number of photos and posts they post online and their privacy settings. In the last part of the questionnaire, the users filled the short version of the NEO-FFM personality characteristics test. 
%TH 9/12/12
%The full questionnaire can be viewed online at: https://nyupolystudy.herokuapp.com/

\subsection{Technical Details}
The questionnaire was hosted online on Heroku and the results were processed using the SPSS software. For correlation calculations, we used the Bi-variate Pearson two-tailed correlation.
%Facebook activity questionnaire, internet usage and pessimism

%Phishing email experiment - how phishing is measured - only people who put in their user name and passwrd are considered phished

%other questions

%\vspace{-3mm}

\subsection{Personality Traits}
We calculated the Five Factor Model personality traits according to the questionnaire. We then used the different personality traits - Openness, Conscientiousness, Extraversion, Agreeableness and Neuroticism - and evaluated their correlation to other test variables.

\subsection{Internet usage, pessimism and addiction}
%correlation between pessimism and phishing
%correlation between questions 19 - probability of stealing a password
We asked a list of questions regarding internet usage and pessimism. The questions were mixed together. Some of the questions were related to internet usage, while the others were related to internet pessimism. The questions related to internet pessimism required the user to assess the likelihood that a negative event will happen to him online (for example, that his password will be stolen).
To evaluate the internet usage, we added the values of all the 'usage' questions for the internet questionnaire section. To evaluate the internet pessimism, we added all the values of the 'pessimism' questions and created one combined value.

We also asked eight questions which correspond to users being preoccupied with the internet, giving a measure of internet addiction. The positive answers to these questions were added to create one variable which correlates to users being addicted to online activity. 

\subsection{Phishing}
In this part of the test, the email addresses provided to us by the students in the questionnaire were used. An email was sent to the users promising an Apple product to the first users to click the link. The email had a few typical characteristics of a phishing email, including the ``from'' field not matching the actual address (which the users would see if they put their mouse on the field). The link also showed a text which did not match the actual link address. In addition, the email  contained spelling mistakes and asked for immediate action, which is typical of phishing emails.

The users that did click on the link were forwarded to a screen that looked like a typical Polytechnic screen. However, the actual html address was:\\ http://alphanext.phpfogapp.com/data\_list/index.php?id=394327.\\
The users who clicked on the login button were then considered to be phished.
To maintain confidentiality, our system only kept the data regarding who was phished but not the actual username and passwords.

Our phishing email was clearly a ``prize scam'' email. The email employed a few psychological techniques, meant to get the users to respond. The email seemed to come from an authority (``CSAW services'', where CSAW is a yearly competition held by the Polytechnic University security group). The email requested an immediate response (which reduces the motivation for thorough consideration and is likely to increase impulsive response). The email also triggered visual processing, by mentioning the prize and 
%TH 9/19/12
that
%mentioning 
the product will be distributed to students (therefore seems ``personalized'' to the University students).

%add figure

A copy of the original phishing email with the phishing characteristics can be found in \figref{phishingemail}.

\begin{figure*}[ht]
  \centering
  \includegraphics[width=0.8\textwidth]{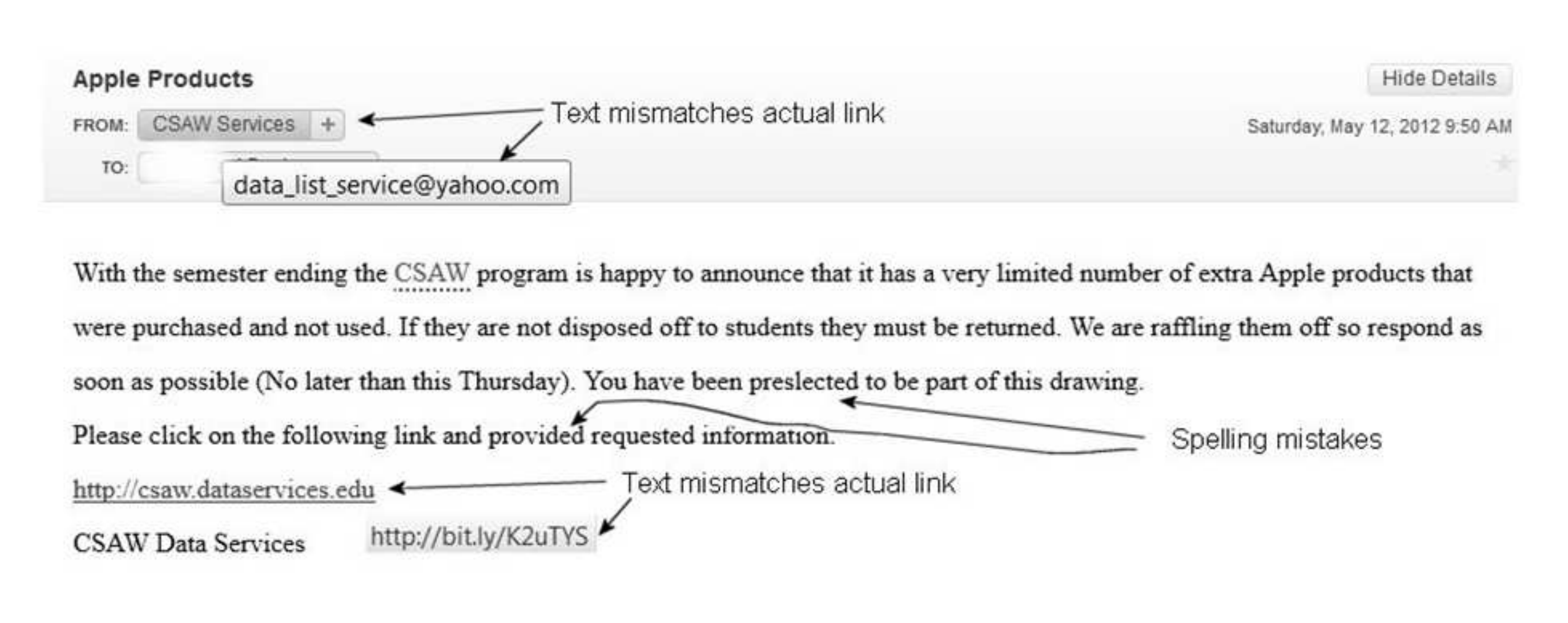}
  \caption{Phishing email}
  \label{phishingemail.fig}
  \vspace{5mm}
\end{figure*}

%\vspace{-3mm}
\subsection{Facebook Activity}
%\vspace{-2.5mm}

To correlate the Facebook activity with the personality traits, we calculated the following: We asked the test participants what kind of data they put on Facebook. The users were asked about 14 different types of personal data, including age, address,   phone number and other personal information.
When examining the entered data, we create variables that reflect the data types the user puts on Facebook - where the variable gets the value $1$ if the user puts the corresponding data on Facebook and $0$ otherwise. We then added the values of all the variables to create
% all the data types the users said they put on Facebook - their birth date, images and other information and create 
one 'Facebook data' combined variable.

We also used the log value of the number of posts and the log value of the number of images the users put on Facebook as separate variables. These give a measure of the amount of activity the user actually engages in on Facebook. 
To calculate the updated variables, we used the following calculation:
$$ \textsf{FB posts} = \log_{10} (\textsf{Total Entries} + 0.001) $$
The same calculation was computed for the total Facebook photos.

To evaluate the privacy settings, we gave a weight of 0 to 3 to each privacy setting, where 0 correlates to allowing nobody to see the related Facebook parameter and 3 correlates to everybody being able to see these. We then added these values for the different parameters to create a combined variable for the Facebook privacy settings.

\section{Results and Discussion}  
\label{sec:results}
Our results showed all 100 test participants filled the questionnaire. Some of the students filled the questionnaire twice. All duplicate entries were removed from the database. 
%We processed the rest of the results for the rest of the students using the SPSS 16.0 software.

\subsection{Phishing}

Our experiment showed that $17\%$ of our users were phished. The most obvious parameter correlated to the phishing was gender. Out of our test participants, $14\%$ of the men were phished and $53\%$ of the women were phished.
%TH 9/18/12
While a similar trend was found in prior research \cite {SHK10}, our results
show a significantly higher difference between the percentage of women and men phished.
%Further, since previous 
Previous research \cite{SHOP} found that women tend to use text messages more as well as shop online more. This further demonstrates the fact that women feel more comfortable with digital communication and may be more inclined to reply to emails with commercial offers or prizes.
%TH 9/17/12
Since our email was a prize offer, this may further contribute to the large difference in response to our phishing email between women and men.

We further tested the correlation between personality traits and being phished. For the women, we found a very high correlation to neuroticism.
%TH 9/18/12
For the men, there was no correlation to any personality trait. 
The full results which show the correlation for the women  
%who participated in our experiment 
can be found in Table \ref{Phishedusagewomen}.
We hypothesize that one of the reasons for this difference may be that women feel more comfortable admitting fears (which many of the questions used to measure neuroticism are related to). 
These results seem to support the hypothesis that women are more sensitive to their emotional needs and tend to believe the internet may have the ability to fill those needs. That fact, together with the fact that the email was a prize phishing email, seem to provide a combination that may make women significantly more susceptible to phishing attacks.

%When examining the results for women, there is also a high correlation between being Phished and openness. On the other hand, there is an inverse correlation between being phished and extraversion.

%One reason for the fact that women were phished more may be that women tend to be more communicative.

%add table

%add table
\begin{table*}[ht]
%[h]
\begin{center}
%\begin{minipage}{10cm}
\begin{tabular}{|c|c|c|c|c|}
\hline
&  \textbf{Phished} & \textbf{Usage}  &  \textbf{Pessimism} & \textbf{Addiction}\\
\hline
\hline
\textbf{Neuroticism} & .501*	& -.161	& -.308	& .464\\
\hline
\textbf{Extraversion} & -.330	& .064	& .013	& -.282\\
\hline
\textbf{Openness} & .357	& .090	& .164	& -.173\\
\hline
\textbf{Agreeableness} & -.057	& -.424	& -.226	& -.071\\
\hline
\textbf{Conscientiousness} & -.034	& .220	* & .187	& -.630**\\
\hline
\textbf{Usage}  & .177	& 1	& .828**	& .009\\
\hline
\textbf{Pessimism} & .148	& .828**	& 1	& .054\\
\hline
\textbf{Addiction} & .043	& .009	& .054	& 1\\
\hline

\end{tabular}

\vspace{5mm}
* - Correlation is significant at the 0.05 level (2-tailed).\\
** - Correlation is significant at the 0.01 level (2-tailed).\\

\caption{Phishing and personality factors correlation for women}
\label{Phishedusagewomen}
%\end{minipage}
\end{center}
%\vspace{-1mm}
\vspace{5mm}
\end{table*}

Another question we are attempting to examine is: Can we predict the probability of being phished based on Facebook activity? To examine this, we ran a linear regression test, trying to predict Phishing vulnerability based on the four Facebook variables: Types of data posted, number of posts, number of photos and privacy settings. Our results appear in Table \ref{PhishFBTablewomen}.
This result point to the fact that there is a correlation between Facebook activity and phishing response. This indicates that 
being more active in online social networks
%people who are more active online are likely to be more susceptible 
may cause higher susceptibility
to such attacks. Therefore, people who feel more comfortable with online communication and expressing themselves online may also be more likely to respond to phishing emails. 

%\begin{tabular}{|c|c|c|c|}
%\hline
%%&  \textbf{Value} & \textbf{Std. Err.}\\
%&  \textbf{Correlation} & \textbf{R Square} &  \textbf{Std. Err.}\\
%% & \textbf{Privacy settings}\\
%\hline
%\hline
%\textbf{Phished} &  .448 &	0.201 & .531	\\
%\hline

\begin{table}
%[htbp]
%[h]
\begin{center}
%\begin{minipage}{10cm}
\begin{tabular}{|c|c|c|}
\hline
%&  \textbf{Value} & \textbf{Std. Err.}\\
&  \textbf{Correlation} & \textbf{R Square} \\
% & \textbf{Privacy settings}\\
\hline
\hline
\textbf{Phished} &  .448 &	0.201 	\\
\hline

\end{tabular}

\vspace{5mm}
%* - Correlation is significant at the 0.05 level (2-tailed).\\
%** - Correlation is significant at the 0.01 level (2-tailed).\\

\caption{Linear regression of phishing and Facebook activity (women)}
\label{PhishFBTablewomen}
%\end{minipage}
\end{center}
%\vspace{-1mm}
\end{table}

\subsubsection{Predicting Phishing}

We also found that people are not good at estimating their vulnerability to internet attack. One of the questions we asked our test subject is how do they rate the likelihood of their passwords being stolen. When correlating their responses to the people who were phished, we found the answers were uncorrelated. Further, we only see a low correlation between general internet pessimism and the likelihood of being phished. This further shows that people are not fully aware of the potential internet threats and their ability to avoid phishing attacks.

%TH 8/21/12
We also asked the users about their computer expertise. We found that there is no correlation between general computer expertise and the ability to detect email attacks.
The correlations can be seen in Table \ref{Phishedpessimism}.

This finding suggests that to understand phishing susceptibility, it is preferable to conduct studies in which the users are being phished (vs. asking people to look at phishing emails and detect that look suspicious). It raises the likelihood that the susceptibility of people to phishing results from failing to consider the possibility that an email may be phishing, but rather concentrating on the potential for gain (prize). 

\begin{table}[htbp]
%[htbp]
%[h]
\begin{center}
%\begin{minipage}{10cm}
\begin{tabular}{|c|c|c|c|c|}
\hline
&  \textbf{Phished} & \textbf{Pessimism}  &  \textbf{Est. Risk} & \textbf{Expert.}\\
\hline
\hline
\textbf{Pessimism}  & .135	& 1	& .725** & -0.54\\
\hline
\textbf{Est. Risk} & -.029 &	.725**	 &1 & 0.100\\
\hline
\textbf{Expertise} & -.044 &	-.054	 &0.100 & 1\\
\hline

\end{tabular}

\vspace{5mm}
* - Correlation is significant at the 0.05 level (2-tailed).\\
** - Correlation is significant at the 0.01 level (2-tailed).\\

\caption{Phishing results correlated to Pessimism and estimated risk (all test participants)}
\label{Phishedpessimism}
%\end{minipage}
\end{center}
%\vspace{-1mm}
\end{table}

%todo - add table

%\newpage

\subsection{Internet usage, pessimism and addiction}
We found that people who use the internet more are also more aware of its risks. They regarded the likelihood of something bad happening to them online higher than the people who use it less. This tends to show that people who spend more time online do become aware of the fact that the internet poses threats to user privacy.
We also found that internet addiction was highly correlated to neuroticism: ($Correlation = 0.426$). 
This is intuitive as people with high neuroticism level tend to become more vulnerable to different addictions. 
%TH 9/17/12
We further see that internet addiction is inversely correlated to conscientiousness. This is similar to correlations found in previous study for substance abuse addiction \cite{KN07}.
%which is inline with previous findings for substance abuse addiction
This demonstrates that people who are likely to be vulnerable to other addictions may also be vulnerable to internet addiction, which may be experienced as a safe activity that provides relief from stress. The correlation can be seen in Table \ref{InternetOCEAN}.

%TH 9/17/12
%We further see that internet addiction is inversely correlated to conscientiousness. This is inline with previous findings for substance abuse addiction and supports the hypothesis that people who are vulnerable to 

\begin{table}
%[htbp]
%[h]
\begin{center}
%\begin{minipage}{10cm}
\begin{tabular}{|c|c|c|c|}
\hline
&  \textbf{Usage} & \textbf{Pessimism} & \textbf{Addiction} \\
\hline
\hline
\textbf{Neuroticism} &  .009	& .180	& .426**\\
\hline
\textbf{Extraversion} & .116	& -.048	& -.043\\
\hline
\textbf{Openness} & -.019	& .004	& .055  \\
\hline
\textbf{Agreeableness} & -.053	& -.111	& -.042\\
\hline
\textbf{Conscientiousness} &.186 &	.025	& -.241*\\
\hline
\textbf{Usage} & 1 &	.684** & -.009\\
\hline
\textbf{Pessimism} & .684**	& 1	& .078\\
\hline
\textbf{Addiction} & -.009 &	.078 &	1\\
\hline

\end{tabular}

\vspace{5mm}
* - Correlation is significant at the 0.05 level (2-tailed).\\
** - Correlation is significant at the 0.01 level (2-tailed).\\

\caption{Internet behavior correlation to personality factors}
\label{InternetOCEAN}
%\end{minipage}
\end{center}
%\vspace{-1mm}
\end{table}

We also examined the correlation between internet behavior and Facebook activities. As expected, we found that people who use the internet more also tend to use Facebook more, posting more data and photos on it. People who are more pessimistic about the internet and estimate its risks higher were found on average to post more messages as well as photos to Facebook. This supports the hypothesis that people who actually use the internet more are more aware of its dangers. 
In addition, participants
%People 
who are more preoccupied with the internet (rate higher on the addiction scale) also tend to put more data on Facebook.

%add table
\begin{table}[htbp]
%[h]
\begin{center}
%\begin{minipage}{10cm}
\begin{tabular}{|c|c|c|c|}
\hline
&  \textbf{Usage} & \textbf{Pessimism} & \textbf{Addiction} \\
\hline
\hline
\textbf{FB Data} & .160 &	.160 &	.203*\\
\hline
\textbf{FB photos} & .234*	& .199*	& .094\\
\hline
\textbf{Total Posts} & .241* &	.162	& .062 \\
\hline
\textbf{Privacy Settings} & .072 &	.072	& .122\\
\hline

\end{tabular}

\vspace{5mm}
* - Correlation is significant at the 0.05 level (2-tailed).\\
** - Correlation is significant at the 0.01 level (2-tailed).\\

\caption{Internet behavior correlation to FB activity}
\label{InternetFB}
%\end{minipage}
\end{center}
%\vspace{-1mm}
\end{table}

%\newpage

\subsection{Facebook Activity}

We also examine the Facebook activity correlation to personality traits.
As expected, we found that Facebook activity correlates to openness, which was correlated with both the data types the users put on Facebook as well as the number of posts and images. Also, openness was correlated with looser Facebook privacy settings. Our tests did not show significant difference between the Facebook activity of men and women. 
Another observation was that Facebook activity is directly correlated to the Facebook privacy settings -  people who are more active on Facebook also tend to have looser privacy settings (less restricted). The full results can be seen in Table \ref{FBTable}. 
%TH 9/18/12
These results indicate people who put more information on Facebook have significantly higher risk of privacy leaks, as they also tend to share this information with significantly more people. This suggests Facebook users who enjoy using the application fail to consider its privacy leak risks while focusing mainly on its advantages.

%Following are notes for Table \ref{FBTable}: \\
%** - Correlation is significant at the 0.01 level (2-tailed).\\
%* - Correlation is significant at the 0.05 level (2-tailed).\\

\begin{table*}[htbp]
%[htbp]
%[h]
\begin{center}
%\begin{minipage}{10cm}
\begin{tabular}{|c|c|c|c|c|}
\hline
&  \textbf{FB Data} & \textbf{FB photos} & \textbf{FB Posts} & \textbf{Privacy}\\
% & \textbf{Privacy settings}\\
\hline
\hline
\textbf{Neuroticism} &  .103 &	.017	&.108 &	.105\\
\hline
\textbf{Extraversion} & .182 & .191	& .134 &	.093\\
\hline
\textbf{Openness} & .306**	& .249*	 &.155	& .251*  \\
\hline
\textbf{Agreeableness} & .005	& .081 &	.096 &	.111\\
\hline
\textbf{Conscientiousness} & -0.003 &.187 &	.116 &	.046\\
\hline
\textbf{FB Data} & 1 &	.744** & .659** &	.696**\\
\hline
\textbf{FB photos} &.744** &	1	& .774** &	.763**\\
\hline
\textbf{FB Posts} & .659** &	.774**	& 1	& .723**\\
\hline
%\textbf{Privacy settings} & .696**	& .763**	& .723**	& 1\\
\textbf{Privacy} & .696**	& .763**	& .723**	& 1\\
\hline

\end{tabular}

\vspace{5mm}
* - Correlation is significant at the 0.05 level (2-tailed).\\
** - Correlation is significant at the 0.01 level (2-tailed).\\

\caption{Facebook data correlation to personality factors}
\label{FBTable}
%\end{minipage}
\end{center}
%\vspace{-1mm}
\end{table*}

\subsection{Users without Facebook accounts}
Within our test population, we found that a small group of 
%We found that was a small group of 
12 test objects had no Facebook account. Inspection of the group showed they were all men and none of them were phished. 
Examining the Five Factor Model variables, we found that the highest inverse correlation for people in this group was to openness while there was also a lower inverse correlation to extraversion. The correlation between the non-Facebook users and the personality traits can be seen in Table \ref{NoFBUsers}.

The results suggest there are certain participants that manifest their off-line personal traits (scoring lower on openness and extraversion) in their online activity as well and are not interested in social networks. This further suggests that people who do not feel comfortable with social online activity may also be less likely to fall victims to online phishing attacks.
%people in the group had a lower measure of Openness (average 21.67 vs. 26.89 for the whole group). 

%add table
\begin{table}
%[htbp]
%[h]
\begin{center}
%\begin{minipage}{10cm}
\begin{tabular}{|c|c|}
\hline
&  \textbf{No FB account} \\
\hline
\hline
\textbf{Neuroticism} & -.070\\		
\hline
\textbf{Extraversion} & -.170\\
\hline
\textbf{Openness} & -.301**\\
\hline
\textbf{Agreeableness} & -.118\\
\hline
\textbf{Conscientiousness} & -.127\\
\hline

\end{tabular}

\vspace{5mm}
* - Correlation is significant at the 0.05 level (2-tailed).\\
** - Correlation is significant at the 0.01 level (2-tailed).\\

\caption{Correlation between user with no Facebook account and personality factors}
\label{NoFBUsers}
%\end{minipage}
\end{center}
%\vspace{-1mm}
\end{table}

\section{Conclusions and Future Work}
\label{sec:conclusions}

Our research examines the factors that may contribute to susceptibility to online security and privacy attacks. We look at the correlation between personality traits and phishing email response. We further examine the correlation between online behavior and probability of being phished.

Our findings have important implications, as they confirm that certain personality traits may cause higher phishing vulnerability. Specifically, we found that women tend to be more susceptible to prize phishing attacks than men. In particular, we saw a high correlation between neurosis and responding to phishing attacks. This suggests phishing defenses should be tailored towards people who score high on certain personality traits, specifically in cases of phishing emails that seem to offer financial gain (such as prizes).

We also see that Facebook activity can be a predictor of vulnerability to phishing. This can be useful in designing defenses 
%designed 
for specific demographics (for example, a defense may be designed as a Facebook application).

Our work also finds that people who are more engaged with Facebook activity (posting more messages and photos) also have less restrictive privacy settings and therefore may be more vulnerable to privacy threats. 
%TH 9/18/12
This suggests people who focus more on the benefits of Facebook tend to ignore its risks, a factor that should be considered when attempting to raise awareness about privacy leaks through user education.
% (as a larger segment of Facebook users get to see their personal information). 

Future work should concentrate on email phishing attacks with different email types. The emotional motivations for responding to different email types may be different. Therefore, finding which personality factors are correlated to the different types will be useful in future design of defenses for online attacks.

%women more communicative?

%women more honest than men about being neurotic

%Results: Correlation to variables - OCEAN, FB data, FB posts, FB pictures, FB privacy settings, Gender

%correlation results

%specific group of without any facebook account

%The authors used spectrum analysis to detect the beginning of each key before
%classifying it.

\section{Acknowledgments}
\label{sec:ack}

This work was supported in part by the NSF (under grant 0966187). The views and conclusions contained in this document are those of the authors and should not be interpreted as necessarily representing the official policies, either expressed or implied, of any of the sponsors.

%
%%\begin{floatingfigure}[r]{.35\textwidth}
%\begin{figure}
%%\begin{minipage}[b]{.45\linewidth}
%\begin{center}
%\vspace{-3mm}
%\includegraphics[width=7cm,height=4cm]{images/signal3ftm.pdf}
%%TH
%\vspace{-5mm}
%\caption{\footnotesize Audio signal for the full key}
%\label{signal3ftm.fig}
%\end{center}
%%\end{floatingfigure}
%%\end{minipage}
%%\vspace{-5mm}
%\end{figure}

%\begin{figure}
%%\hspace{.2cm}
%%\begin{minipage}[b]{.45\linewidth}
%\begin{center}
%%\flushleft
%\vspace{-6mm}
%{\includegraphics[width=7cm,height=4cm]{images/onezerosignal3ftm.pdf}}
%\vspace{-5mm}
%\caption{\footnotesize Acoustic signal (in meat)}
%\vspace{-7mm}
%\label{onezerosignal3ftm.fig}
%\end{center}
%%\end{minipage}
%%TH 7/21/10
%%\vspace{-7mm}
%%\vspace{-5mm}
%\end{figure}

%{\footnotesize{
\bibliographystyle{ieee}
\bibliography{paper}
%}}

\end{document}